\documentclass[final,5p,times,twocolumn,english]{elsarticle}

\makeatletter
\providecommand{\tabularnewline}{\\}

\makeatother

\usepackage{babel}

\usepackage{epsfig}
\usepackage{array}
\usepackage{graphicx}

\newcommand{\beq}{\begin{equation}}
\newcommand{\eeq}{\end{equation}}

\journal{Physics Letters A}

\begin{document}
 \begin{frontmatter}
\title{Non-Gaussian buoyancy statistics in fingering convection}

\author[1]{Jost von Hardenberg}
\ead{jost@isac.cnr.it}
\author[2]{Francesco Paparella}
\ead{francesco.paparella@unisalento.it}

\address[1]{Institute of Atmospheric Sciences and Climate - CNR, Torino, Italy }
\address[2]{Dip. di Matematica-Universit\'a del Salento and CMCC, Lecce, Italy}

\begin{abstract}
  We examine the statistics of active scalar fluctuations in 
  high-Rayleigh number fingering convection with high-resolution
  three-dimensional numerical experiments. The one-point 
  distribution of buoyancy fluctuations
  is found to present significantly non-Gaussian tails. 
  A modified theory based on an original approach by Yakhot (1989) 
  is used to model the active scalar distributions 
  as a function of the conditional expectation values of scalar 
  dissipation and fluxes in the flow. 
  Simple models for these two quantities
  highlight the role of blob-like coherent 
  structures for scalar statistics in fingering convection.
\end{abstract}
\begin{keyword}
Doubly-diffusive convection; Salt fingers 
\end{keyword}

\end{frontmatter}

\section{Introduction}

Fingering convection is a peculiar convective flow, characterized by a
counter-gradient density transport, which is of interest for a wide
range of fields, including stellar physics, metallurgy and
volcanology\citet{Turner74}. It raises particular interest in
oceanography\citet{Schmitt94,Schmitt03}, where warm, salty, and light
waters floating above fresher, colder and denser waters create
finger-favorable conditions. This situation is commonplace in the
thermocline of subtropical oceans, where finger-generated diapycnal
mixing is believed to affect the meridional overturning circulation,
and the uptake of heat and carbon dioxide \citet{Schmitt05}. Fingering
convection occurs when two buoyancy-changing scalars with different
diffusivities are stratified in such a way that the least-diffusing
one, if taken alone, would produce an upward, unstable density
gradient, but the most-diffusing one reverses this tendency and
produces a net downward density gradient. This set-up allows for a
doubly-diffusive instability, where infinitesimal perturbations to the
initial stratification may undergo exponential growth. If a fluid
parcel is displaced downward it looses the stabilizing, most-diffusing
scalar at a faster rate than the destabilizing, least-diffusing one,
because of the difference in diffusivities. This results in an
increase of the density of the parcel, which sinks at a lower depth,
where it loses even more stabilizing scalar. A symmetrical argument
holds for a fluid parcel displaced upward. While the two scalars are
transported along their gradients, a net counter-gradient buoyancy
flux results from this mechanism.  In the linear theory maximum
growth rate is attained at horizontal wavelengths sufficiently small so
that the difference in diffusivity of the two scalars is important,
but not so small that the damping effect of viscosity becomes dominant
(a few centimeters in the oceans), and at a vertical wavelength
corresponding to the vertical domain extension
\citet{Baines&Gill69,Stern75}. In the fully nonlinear regime, the
elongated, finger-like columns of the linear theory break down into
shorter plumes, or blobs, as we shall call them, of rising and
sinking fluid, having an aspect ratio much closer to one. Vigorous
convection arises, dominated by the complicated dynamics of the
interacting blobs \citet{Merryfield00,Radko08}.  In some instances a
secondary instability disrupts the linear profiles of horizontally
averaged temperature and salinity, leading to the formation of the
so-called ``staircases'' where finger zones are alternated with slabs
of vertically well-mixed fluid \citet{Schmitt94,Krishnamurti03}.

While ample literature has been devoted to convection problems with a
single scalar, the Rayleigh-B\'enard set-up being the leading example,
convection with two active scalars, such as fingering convection, has
yet to be explored to the same extent. In particular, knowledge on the
statistical properties of scalar fluctuations can be useful for
constraining and testing models of the convective scalar fluxes.
In this paper we use high-resolution three-dimensional numerical
simulations to explore the high-Rayleigh number regime of fingering
convection and we analyze and interpret the scalar fluctuation
distributions in these flows using a novel formulation of a classical
theoretical approach.

\section{The Simulations}

\subsection{The equations of fingering convection}

As customary in this problem, we denote the least-diffusing scalar as
salinity, $S$, and the most-diffusing one as temperature, $T$, even if
the actual physical nature of the two scalars may be different. We
consider a fluid layer of thickness $d$ confined above and below by
perfectly conducting, parallel, plane plates mantained at constant
temperature and salinity. We bring the problem to a non-dimensional
form by scaling temperatures and salinity with their plate differences
$\Delta T$ and $\Delta S$, scaling lengths with the layer thickness,
$d$, and using the haline diffusive time, $\tau_S=d^{2}/\kappa_{S}$,
as a timescale, with $\kappa_{S}$ the haline diffusivity. The control
parameters of the problem are the Prandtl, Lewis, thermal and haline
Rayleigh numbers, defined as
\[
Pr=\frac{\nu}{\kappa_{T}},\, Le=\frac{\kappa_{T}}{\kappa_{S}},\, R_{T}=\frac{g\alpha\Delta T\, d^{3}}{\nu\kappa_{T}},\, R_{S}=\frac{g\beta\Delta S\, d^{3}}{\nu\kappa_{S}},
\]
respectively. In these expressions $\nu$ is the kinematic viscosity,
$\kappa_{T}$ is the thermal diffusivity, $g$ is the modulus of the
gravity acceleration and $\alpha$ and $\beta$ are the thermal and
haline linear expansion coefficients. From these the \emph{density
  ratio} $R_{\rho}=Le\, R_{T}/R_{S}$ can be defined. Linear stability
analysis shows that a necessary condition for fingering instability is
$1<R_{\rho}<Le.$ The density ratio controls directly the intensity of
convection in the flow. Values close to one lead to fast instability
and violent convection.  The haline diffusive time is the longest time
scale in this problem. Far from marginality, it is often more
convenient to use the convective time scale
$\tau_c=(Pr\,Le\,R_{S})^{-1/2}\tau_S$.

In the Boussinesq approximation, the resulting non-dimensional equations
for fingering convection are: 
\begin{eqnarray}
\frac{\partial\mathbf{u}}{\partial t}+\mathbf{u}\cdot\nabla\mathbf{u} & = & -\nabla p+Pr\, Le\left[R_{S}B\,\hat{\mathbf{z}}+\nabla^{2}\mathbf{u}\right],\label{eq:Ueq}\\
\frac{\partial T}{\partial t}+\mathbf{u}\cdot\nabla T & = & Le\,\nabla^{2}T\label{eq:Teq},\\
\frac{\partial S}{\partial t}+\mathbf{u}\cdot\nabla S & = & \hphantom{Le\,}\nabla^{2}S\label{eq:Seq},\\
\nabla\cdot\mathbf{u} & = & 0 ,\label{eq:continuity}
\end{eqnarray}
where $\mathbf{u}=(u,v,w)$ is the solenoidal velocity field of the
fluid, $p$ is the pressure, and we have defined the \emph{buoyancy}
field $B=R_{\rho}T-S$. The latter is the dynamically important linear
combination of the $T,$ $S$ scalar fields as it appears in the forcing
term of the momentum equation (\ref{eq:Ueq}).  Notice that the
equations could be rewritten entirely in terms of buoyancy together
with some other linear combination of temperature and salinity. In
this case buoyancy would play the role of the active,
momentum-generating scalar, and the other scalar could be considered
almost passive, interacting only with buoyancy through a diffusive
term.  For example in the oceanographic literature a quantity called
``spice'' is sometimes defined \citet{Flament02} as the linear
combination of temperature and salinity which is maximally independent
of buoyancy.

We integrate the equations numerically with a code which is
pseudospectral in the horizontal directions and with finite
differences in the vertical, with a non-homogeneous vertical grid in
order to better resolve the boundary layers and with a third-order
fractional step method for time advancement
\citet{Passoni02,Parodi04,Parodi08}. Free slip boundary conditions are
used at the top and at the bottom and a laterally periodic domain is
assumed.

While the density ratio represents the main control parameter in this
type of flows, in the sense that small variations of $R_{\rho}$
determine large variations in the fluxes, we study the changes
in the flow determined by changes in the vertical extent of the
domain, as encoded in the Rayleigh numbers.  It has been suggested
that simulations in a vertically periodic domain should yield results
comparable with those that could be observed in a physical situation
where the vertical extension is many orders of magnitude larger than
the scale of the fingers (e.g. \citet{Merryfield00,Radko08}).
However, recent work \citet{Calzavarini06} casts shadows on the
meaningfulness of investigations in such a domain. This requires
treating the magnitude of the Rayleigh numbers, and not only their
ratio, as an independent free control parameter of the problem.
Accordingly here we use a configuration confined between upper and
lower rigid plates, and push up the Rayleigh numbers while maintaining
a fixed density ratio $R_{\rho}=1.2$, which is sufficient to guarantee
vigorous convection.  We perform a series of numerical experiments,
fixing $Pr=10, Le=3$, 
and exploring the range $R_{S}=10^{8}-10^{11}$.  The vertical domain
extension is $-0.5<z<0.5$. The vertical resolution ranges between
$N_{z}=145$ (at $R_{S}=10^{8}$) and $N_{z}=809$ layers (at
$R_{S}=10^{11}$).  The horizontal resolution is maintained at
$N_{x}=192$ grid points, while the lateral domain size is scaled in
order to be approximately proportional to the horizontal linear
instability scale \citet{Stern75}, according to
$L=(R_{S}/10^{9})^{-1/4}$.

\subsection{Statistics of scalar fluctuations}

\begin{figure}
\begin{centering}
\includegraphics[width=0.5\textwidth]{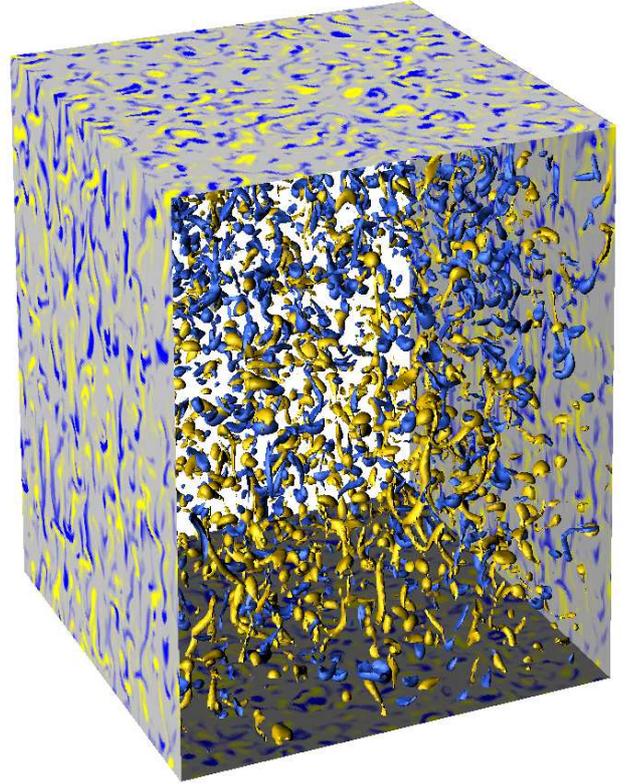}
\par\end{centering}
\caption{\label{fig:render} Rendering of buoyancy fluctuation
  iso-surfaces at the level $2.5\sigma_B$, once statistical
  stationarity has been reached, in the center of the domain between
  $z=-0.2$ and $z=0.2$.  The lateral panels show sections through the
  flow with a colormap saturating at $3\sigma_B$.  Yellow(blue)
  corresponds to positive(negative) buoyancy fluctuations. }
\end{figure}

\begin{figure}
\begin{centering}
\includegraphics[width=0.46\textwidth]{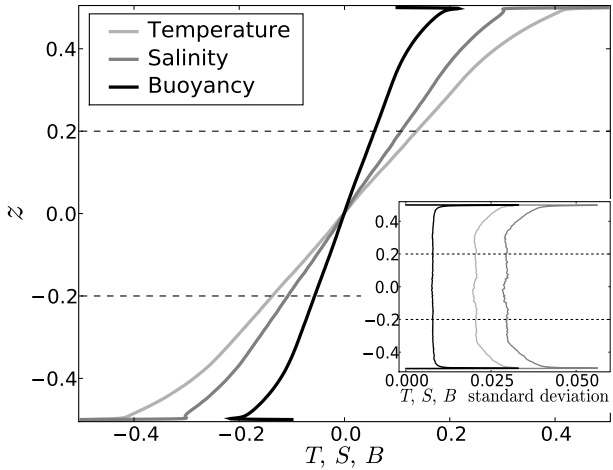}
\par\end{centering}
\caption{\label{fig:profs} Horizontally and time averaged scalar
  profiles for the simulation at $R_{S}=10^{11}$; (in the inset)
  profiles of the standard deviation of scalar fluctuations at fixed
  $z$. The time averages extend over one hundred convective time
  units.}
\end{figure}

Fig. \ref{fig:render} shows an isosurface rendering of the buoyancy
fluctuation field for the simulation at $R_{S}=10^{11}$, in the center
of the domain, after the solution has reached statistical
stationarity. The flow is characterized by the presence of well
defined buoyancy structures which transport a large fraction of the
vertical buoyancy fluxes and lead to the vertical homogenization of
the statistical properties of the flow\footnote{For example in the
  data in Fig. \ref{fig:render}, points with buoyancy fluctuation
  larger than 2 standard deviations in modulus cover only 5\% of the
  volume but carry 37\% of the total vertical buoyancy flux.}.
Fig. \ref{fig:profs} shows vertical profiles of horizontal averages of
buoyancy, temperature and salinity, and of the variance of
fluctuations respect to these profiles. Note the inverse boundary
layers of buoyancy, which testify of a counter-gradient advective
transport. An ample central region characterized by a uniform linear
background and approximately $z-$independent variance is evident.
This is the region of interest for many geophysical applications of
fingering convection, including oceanographic problems, all
characterized by high Rayleigh numbers and reduced influence of the
vertical boundary conditions. Increasing the Rayleigh numbers in our
simulations leads to a growth of the height of this central region.
The presence of a vertical average background gradient of the scalars
is characteristic of fingering convection, and it is consistent with
laboratory experiments and observations (e.g
\citet{Schmitt94,Krishnamurti03}) where approximately uniform vertical
mean gradients are reported within the fingers zones.

In the central region buoyancy fluctuations (respect to horizontally
averaged buoyancy), vertical velocities, and convective fluxes all vary according
to approximate power laws with the Rayleigh number, as summarized in
table \ref{tab:statistics}. The table also reports two independent
estimates of the characteristic scales of the structures in the
buoyancy field: $l_{x}$ is the position of the first zero of the
spatial autocorrelation function of buoyancy, computed keeping
constant the coordinates $y$ and $z$; $l_{v}$ is the cubic root of the
average volume of the connected regions with
$|B^{\prime}|\ge2\sigma_{B^{\prime}}$. Both estimates give similar
results.

\begin{table}
\begin{centering}
\footnotesize
\begin{tabular}{|c|c|c|c|c|c|c|}
\hline 
$R_S$ & $\sigma_{B'}$  & $\sigma_{W}$  & $\langle WB^{\prime} \rangle$ & $l_{x}$  & $l_{v}$  & $Re$\tabularnewline
\hline 
     
$10^{8}$  & 0.0159 & 432.2 & 5.81 & 0.0402 & 0.0447 & 0.58\tabularnewline
$10^{9}$  & 0.0132  & 1017.6  & 11.06  & 0.0218 & 0.0243 & 0.74\tabularnewline
$10^{10}$  & 0.0099  & 2621.0 & 19.45 & 0.0128 & 0.0138 & 1.12\tabularnewline
$10^{11}$  & 0.0078  & 6271.5 & 33.56  & 0.0076 & 0.0070 & 1.59\tabularnewline
\hline
Exponent:  & -0.11  & 0.39  & 0.25  & -0.24  & -0.26  & \tabularnewline
\hline
\end{tabular}
\par\end{centering}
\caption{\label{tab:statistics} Statistics of buoyancy fluctuations as 
  a function of haline Rayleigh number, computed at the center of the 
  domain ($-0.2\leq z\leq0.2$), once statistical stationarity has been
  reached. The standard deviations of buoyancy fluctuations and of vertical velocities,
  $\sigma_{B^{\prime}}$ and $\sigma_{W}$, and the vertical
  convective fluxes, $\langle {WB^{\prime}} \rangle$, are reported; $l_{x}$ and $l_{v}$ are
  two independent estimates of the size of the structures in the
  buoyancy field (see text for details). The rightmost column reports
  the Reynolds number of the buoyancy structures estimated as
  $Re=\sigma_{W}l_{x}/(LePr)$. The bottom row reports exponents
  obtained fitting a power law scaling respect to $R_{S}$.}
\end{table}

\begin{figure}
\begin{centering}
  a)\includegraphics[clip,width=0.46\textwidth]{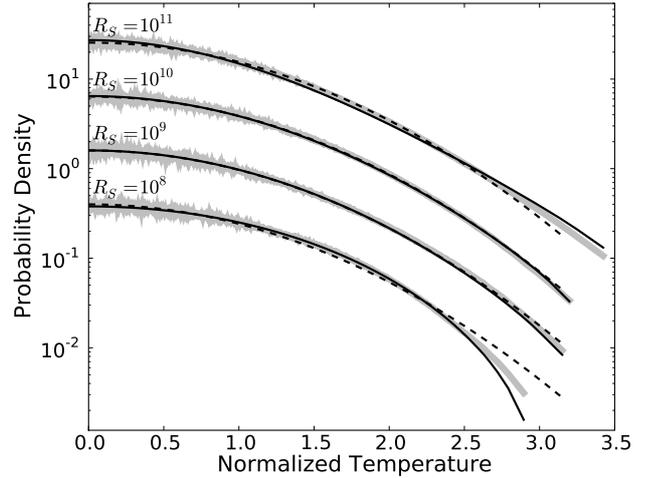} 
  b)\includegraphics[clip,width=0.46\textwidth]{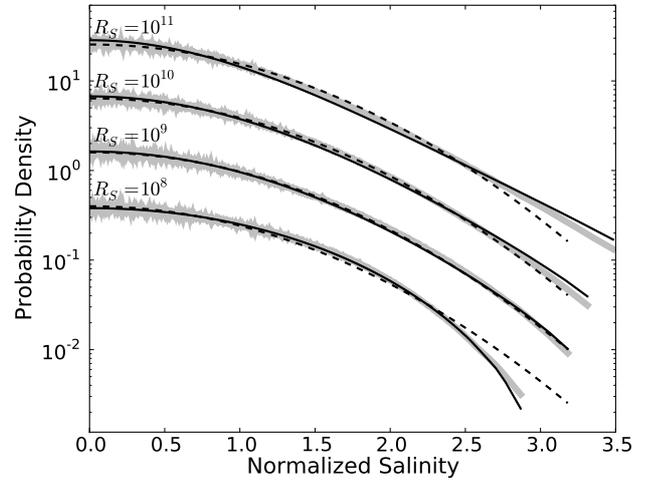}\\
  c)\includegraphics[clip,width=0.46\textwidth]{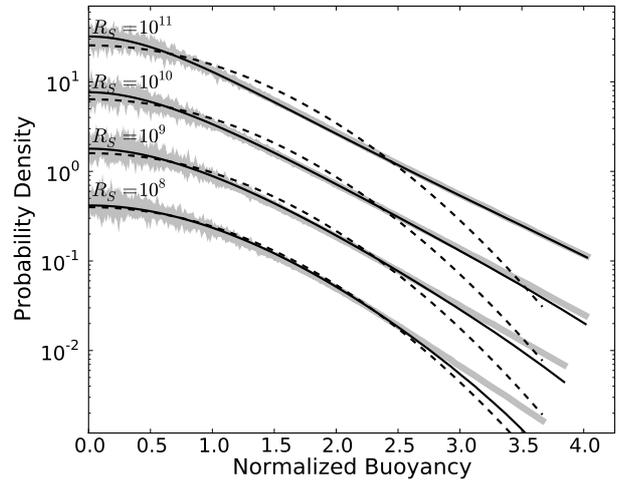}
\par\end{centering}
\caption{\label{fig:pdfsB}$(a,b,c)$ Probability density functions of the 
  temperature, salinity and buoyancy fluctuations respectively, normalized to unitary variance. 
  Data for four different simulations at
  $R_{S}=10^{8}, 10^{9}, 10^{10}, 10^{11}$ are reported, gathered over a
  time interval of one hundred convective times, once stationarity has
  been reached.  For clarity the latter three curves are multiplied by
  two, four and eight units, respectively. Superimposed on the
  numerical data are expression (\ref{eq:P(X)}) and its analogues for temperature and salinity, 
  evaluated using the
  fits (\ref{eq:fitchiscal}) (for temperature and salinity), 
   (\ref{eq:fit-EChiB}) (for buoyancy) and $E(\mathcal{F}|X)/X=X$.
  Gaussian distributions with unitary variance are reported for reference, 
  using dashed lines. }

\end{figure}

Figs. \ref{fig:pdfsB}(a-c) compare the one-point probability
distributions of temperature, salinity and buoyancy fluctuations, for
haline Rayleigh numbers $R_{S}=10^{8},10^{9},10^{10},10^{11}$,
computed at the center of the domain, $-0.2\leq z\leq0.2$, once
statistical stationarity has been reached. Temperature and salinity
present sub-Gaussian distributions at low $R_{S}$, which become
roughly Gaussian at intermediate $R_{S}$ and develop tails longer than
Gaussian only at the highest $R_{S}$ explored in this work. The
distribution of buoyancy, instead, shows non-Gaussian tails
already at the lowest $R_{S}$ considered, which become very similar to
exponentials at high $R_{S}$.

The emergence of exponential tails in the active scalar distribution is
reminiscent of the classical transition to so-called {}``hard
turbulence'' for thermal convection, first reported in
\citet{Castaing89}. In that case buoyancy is determined by only one
scalar (temperature), which develops well defined exponential tails at
high Rayleigh number. The behaviour of a passive scalar
in homogeneous turbulence has been discussed in e.g. in 
\citet{Sinai&Yakhot89,ShraimanSiggia00} and also in that case 
non-Gaussian behavior is found.
The present case is only partially similar to those just mentioned:
no scalar is completely passive in fingering convection, and the buoyant
structures have a very different origin and morphology compared to single-scalar
convection (finger-like blobs vs. turbulent plumes).

\section{A model for the fluctuation distributions}

\subsection{Yakhot's approach}

The scalar probability distributions described above can be
interpreted by adapting to the present case with two active scalars
the theory developed by Yakhot\citet{Yakhot89} for the case of
high-Rayleigh number thermal convection. An important pre-requisite of
that theory is a constant, non-zero vertical gradient of the
horizontally averaged density field, a situation which does not occur
in the bulk flow of Rayleigh-B\'enard convection, but whose presence
is, as we have discussed above, the hallmark of fingering convection.

In order to proceed along these lines we express the buoyancy field as
$B(x,y,z,t)=B^{\prime}(x,y,z,t)+G_{B}z$, where $G_{B}$ is the vertical
gradient of the horizontally averaged buoyancy, and $B^{\prime}$ is
the buoyancy fluctuation. The quantities $T^{\prime}$, $S^{\prime}$,
$G_{T}$ or $G_{S}$ can be defined analogously. Combining
Eq. (\ref{eq:Teq}) and (\ref{eq:Seq}) we obtain an evolution equation
for the buoyancy fluctuation, written in terms of buoyancy and
temperature:
\begin{equation}
  \frac{\partial B^{\prime}}{\partial t}+\mathbf{u}\cdot\nabla B^{\prime}-wG_{B}= R_{\rho}\left(Le-1\right)\nabla^{2}T^{\prime}+\nabla^{2}B^{\prime} .
\end{equation}
We multiply this expression by $B^{\prime2n-1}$ and denote with
$\left\langle \cdot\right\rangle$ a time and volume average. The
volume averages are carried over a central portion of the domain,
where the assumption of constant gradients holds (represented by the
dashed lines of Fig. \ref{fig:profs} for our simulations).  Assuming a
statistically steady state we can integrate by parts; the lateral
boundary terms are zero for periodic (or no flux) boundary conditions;
the top and bottom terms cancel each other because statitical
stationarity implies that the vertical flux of all moments of
$B^{\prime}$ must be the same at all heights. Dividing the resulting
equation for a generic $n$ by the equation for $n=1$ we
obtain
\begin{equation} (2n-1)\left\langle
    X_{B}^{2n-2}\chi_{B}\right\rangle =\left\langle
    X_{B}^{2n-2}\mathcal{F}_{B}\right\rangle , \label{eq:X_variance}
\end{equation}
where we have defined the normalized buoyancy fluctuation,
$X_{B}=B'/\left\langle B'^{2}\right\rangle ^{1/2}$, the normalized
buoyancy flux, $\mathcal{F}_{B}=wB'/\left\langle wB'\right\rangle $
and the normalized buoyancy dissipation rate, $\chi_{B}=\frac{ \nabla
  B^{\prime}\cdot\left[R_{\rho}\left(Le -1\right)\nabla
    T^{\prime}+\nabla B^{\prime}\right]} {\left\langle \nabla
    B^{\prime}\cdot\left[R_{\rho}\left(Le -1\right)\nabla
      T^{\prime}+\nabla B^{\prime}\right]\right\rangle } $.  We refer
to the latter quantity as 'buoyancy dissipation', even if it may
be negative somewhere in the fluid. In fact, the signature of
doubly-diffusive convection is the ability of the second-order derivative terms
in the equations to behave as sources of buoyancy fluctuations, rather
than solely as sinks. But the large-scale, average effect of
these terms remains that of sinks of variance: for $n=1$
Eq. (\ref{eq:X_variance}) is the balance between the advective rate of
extraction of buoyancy variance from the vertical gradients, and its
diffusive dissipation rate at small scales. Note that this process
cannot be described solely in terms of buoyancy: temperature or
salinity fluctuations must appear explicitly in the buoyancy variance
dissipation rate.

In the following we interpret the quantities $X_{B},$
$\mathcal{F}_{B},$ and $\chi_{B}$ as stochastic variables and we make
our last assumption, namely the equivalence of space-time and ensemble
averages. We observe that the probability density of $X_{B}$ must have
a compact support, because Eqs. (\ref{eq:Teq}) and (\ref{eq:Seq})
satisfy a maximum principle, and this forbids arbitrarily large
buoyancy fluctuations; therefore the averages that appear in
Eq. (\ref{eq:X_variance}) exist for any integer $n>0$.

From Eq. (\ref{eq:X_variance}) it is possible to follow Yakhot's
analysis\citet{Yakhot89} slavishly, obtaining an explicit expression
for the probability density of $X_{B}$:
\begin{equation}
P(X_{B})=\frac{E(\chi_{B}|0)P(0)}{E(\chi_{B}|X_{B})}\exp\left[-\int_{0}^{X_{B}}\frac{E(\mathcal{F}_{B}|x)}{xE(\chi_{B}|x)}\, dx\right] , 
\label{eq:P(X)}
\end{equation}
where $E(\cdot|\cdot)$ denotes a conditional average. This is an exact
relationship for the one-point probability density of buoyancy
fluctuations which depends only on two unknown functions, namely the
expected values of the buoyancy fluxes, $E(\mathcal{F}_{B}|X_{B})$,
and of the dissipation, $E(\mathcal{\chi}_{B}|X_{B})$, conditioned on
the buoyancy fluctuations. The constant $P(0)$ is fixed by the
normalization requirement of the density function.  Far from the
physical boundaries it is reasonable to assume that the actual
solution of the Boussinesq equations
(\ref{eq:Ueq}-\ref{eq:continuity}) will have the same up-down simmetry
of the equations themselves: It follows that $P(X_{B})$, and the unknown
conditional expectations $E(\chi_{B}|X_{B})$ and
$E(\mathcal{F}_{B}|X_{B})$, are even functions of
$X_{B}$. Furthermore, $E(\mathcal{F}_{B}|0)=0$, because
$\mathcal{F}_{B}=0$ where $X_{B}=0$.

Theoretical expressions for the distributions of normalized
temperature and salinity fluctuations, or of their linear
combinations, such as spice, can be obtained following the same
approach. This yields expressions which are functionally identical to
Eq. (\ref{eq:P(X)}), but where (e.g. for temperature) the quantities
$X_{T}=T^{\prime}/\left\langle T^{\prime2}\right\rangle ^{1/2}$,
$\mathcal{F}_{T}=wT^{\prime}/\left\langle wT^{\prime}\right\rangle $,
$\chi_{T}=|\nabla T^{\prime}|^{2}/\left\langle |\nabla
  T^{\prime}|^{2}\right\rangle $ appear in place of $X_{B}$,
$\mathcal{F}_{B}$, $\chi_{B}$. In the following we will also use the
unlabeled symbols $X$, $\mathcal{F}$, $\chi$ for, respectively,
normalized fluctuations, fluxes and dissipation of a generic scalar.

\subsection{Scalar fluxes}\label{subsec:fluxes}

\begin{figure}
\begin{centering}
a)\includegraphics[width=0.46\textwidth]{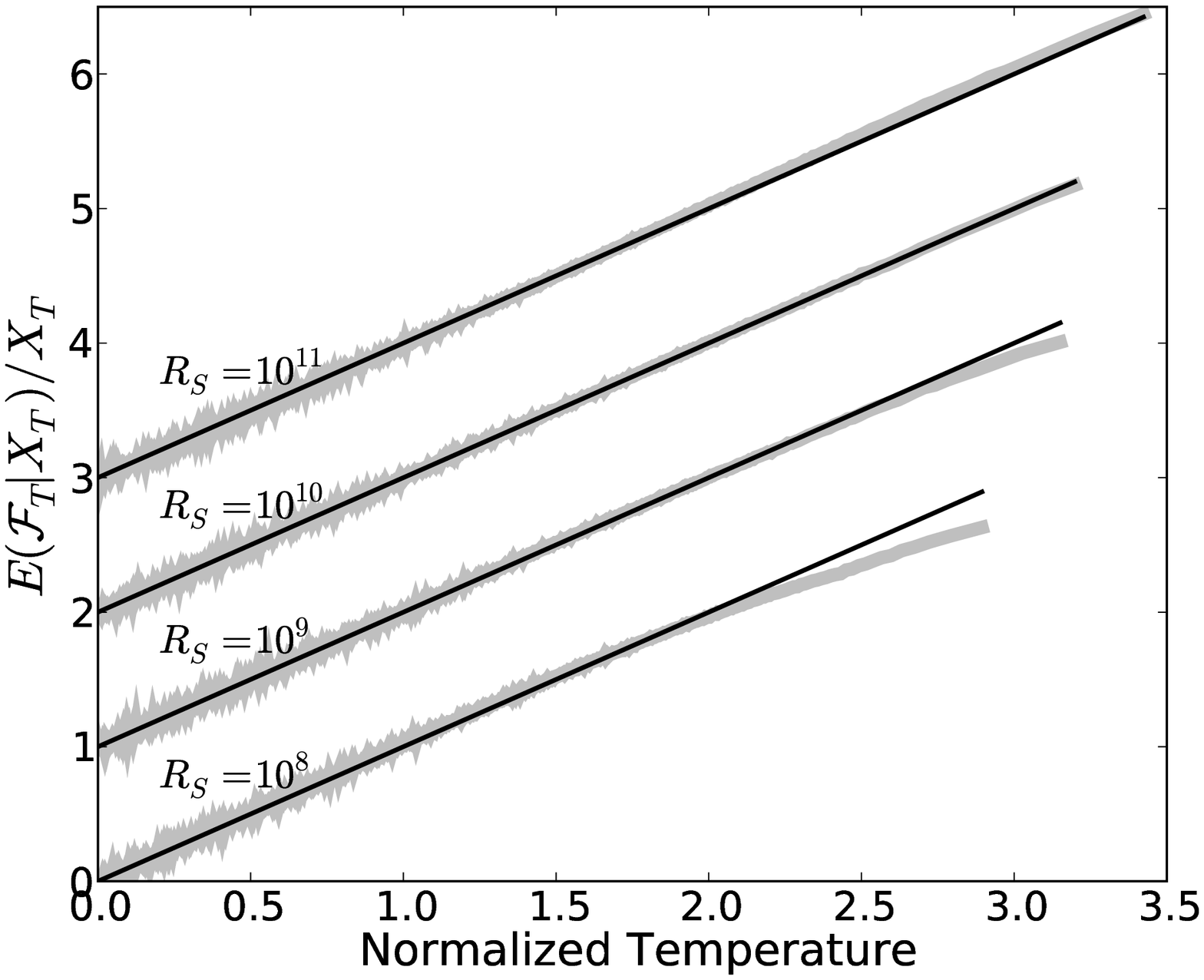}
b)\includegraphics[width=0.46\textwidth]{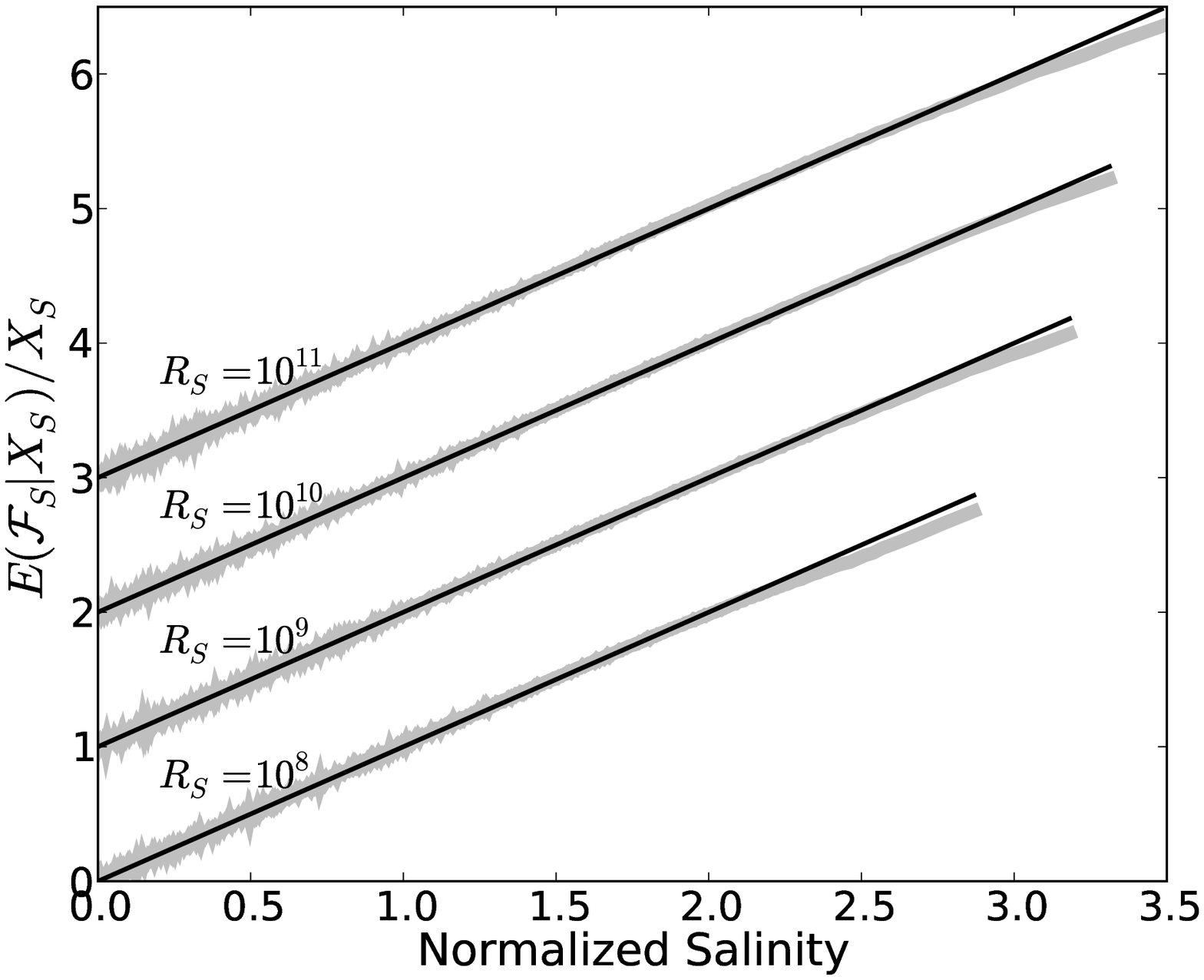}\\
c)\includegraphics[width=0.46\textwidth]{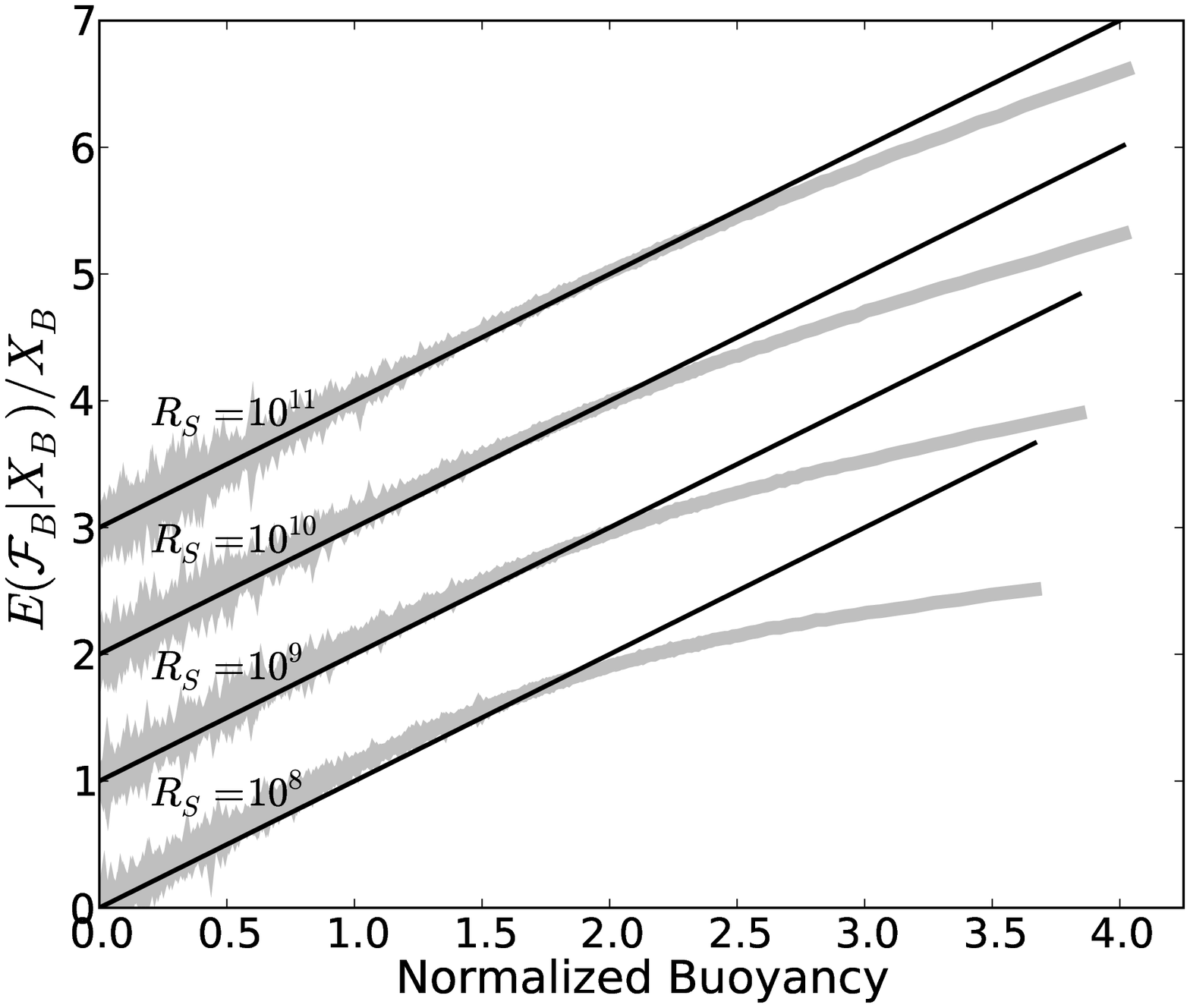}
\par\end{centering}
\caption{\label{fig:Cond-fluxes}$(a)$ Plot of
  $E(\mathcal{F}_{T}|X_{T})/X_{T}$ vs. $X_{T}$ for the four
  simulations at $R_{S}=10^{8}, 10^{9}, 10^{10}, 10^{11}$; data are
  gathered for one hundred convective times, once statistical
  stationarity has been reached.  For clarity the latter three curves
  are shifted upward by one, two and three units, respectively. The
  straight lines are the theoretical fit $E(\mathcal{F}|X)/X=X$.  Only
  the positive normalized fluctuation half-plane is shown. For the
  negative half-plane odd symmetry holds. $(b,c)$ The analogous plots
  for salinity and buoyancy. }
\end{figure}

The normalized, conditional, averages of the temperature and salinity
fluxes in our numerical experiments, $E(\mathcal{F}_{T}|X_{T})/X_{T}$
and $E(\mathcal{F}_{S}|X_{S})/X_{S}$, are reported in
Figs. \ref{fig:Cond-fluxes}(a,b), computed in the positive half-plane.
The linear expression $E(\mathcal{F}|X)/X=X$ fits very well the data,
particularly at larger Rayleigh numbers. Small deviations from
linearity are evident only for large temperature or salinity
fluctuations.  Note that no further constants appear in this
expression due to the identity $\langle \mathcal{F}
\rangle=\int{E(\mathcal{F}|X)P(X) dX}$ and the requirements $\langle
X^2 \rangle=1$ and $\langle \mathcal{F} \rangle=1$.  This result has a
very simple interpretation in the flow under study, if temperature and
salinity are considered as almost passive scalars: as a parcel of
fluid moves downward (upward) in the presence of background gradients,
it carries with it the temperature and salinity values corresponding
to a higher (lower) level, thus generating positive (negative)
temperature and salinity fluctuations. Molecular diffusion tends to
remove the fluctuation, until an equilibrium between these two
competitive factors is found, leading to a fluctuation proportional to
the vertical speed of the parcel of fluid.  From this, the
relationship $E(\mathcal{F}|X)/X=X$ trivially follows.
 
\begin{figure}
\begin{centering}
\includegraphics[width=0.46\textwidth]{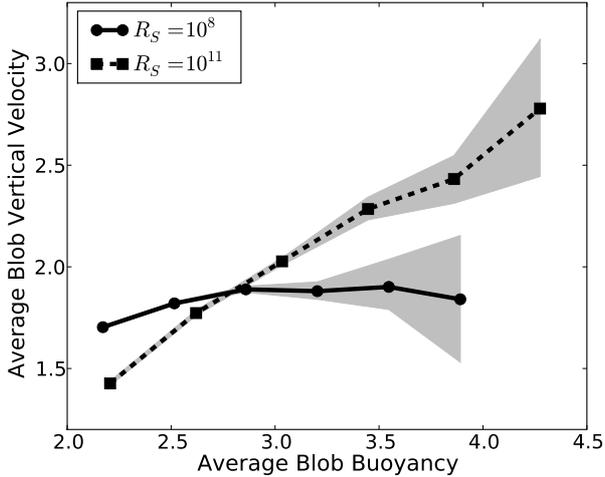}
\par\end{centering}
\caption{\label{fig:WvsB-in-the-blobs} Average vertical velocity vs.
  average buoyancy in connected regions with
  $|B^{\prime}|\ge2\sigma_{B^{\prime}}$ for the $R_S=10^8$ and the
  $R_S=10^{11}$ simulations.  Errobars in gray
  are 95\% bands obtained by jackknife subsampling. Data are gathered
  for one hundred convective times, once statistical stationarity has
  been reached.}
\end{figure}

A different mechanism must apply for buoyancy fluctuations, which play
a direct role in determining vertical accelerations in the flow and
which are generated by the doubly-diffusive mechanism, rather than by
vertical displacement of the fluid.  Fig. \ref{fig:Cond-fluxes}(c)
reports the normalized conditional averages of buoyancy fluxes,
$E(\mathcal{F}_{B}|X_{B})/X_{B}$. While the linear expression
$E(\mathcal{F}_{B}|X_{B})/X_{B}=X_{B}$ again fits the data, more
severe deviations from linearity are evident at large buoyancy values,
particularly at low Rayleigh numbers.  This result is consistent with
a scenario where the advective fluxes are determined to a large extent
by the motion of blobs of buoyancy of characteristic size $l$,
travelling, on average, at a vertical velocity $w$ determined by the
balance between viscous drag and buoyancy forces.  This idea is
supported by the observation that the typical Reynolds numbers of the
structures present in the buoyancy field are low and of order one (see
table 1). If the blobs can be approximated as spheres of diameter $l$,
carrying an average buoyancy fluctuation $B^{\prime}$, the balance
between the buoyancy and the Stokes drag forces would be
$R_{S}B^{\prime}l^{3} \propto lw$. If $l$ is roughly independent of
$B^{\prime}$ we may take $B^{\prime}\propto w$ for uniform blobs. Note
that the power-law exponents for the dependence on $R_{S}$ of
$B^{\prime}$, $w$, $l_{x}$ and $l_{v}$ in Table 1, are in good
agreement with the relationship $w\propto B^{\prime}l^{2}R_{S}$, after
taking the standard deviation of $B^{\prime}$ and $w$ as the
characteristic values for the buoyancy and the vertical velocity
fields.  In this simple scenario the advective buoyancy fluxes are
proportional to the square of the buoyancy fluctuations, leading to
$E(\mathcal{F}_{B}|X_{B})/X_{B}=X_{B}$.  This scenario applies mainly
at high Rayleigh numbers, where the organization of the flow in
approximately spherical blobs is more marked. The internal structure
of the blobs and their interactions will lead to deviations from
linearity, which are apparent in Fig. \ref{fig:Cond-fluxes}(c),
particularly at large buoyancy and at low Rayleigh number.  To have a
further insight into this issue we have partitioned the buoyancy field
into connected regions with $|B^{\prime}|\ge2\sigma_{B^{\prime}}$ and
we have computed the average buoyancy and the average vertical
velocity within each of these regions. The results are shown in
Fig. \ref{fig:WvsB-in-the-blobs}.%
It is evident that at the lowest Rayleigh number the vertical velocity
of these connected regions is fairly independent of the region's
average buoyancy, thus breaking the proportionality $w\propto
B^{\prime}$ for extreme values of $B^{\prime}$. At high Rayleigh
number connected regions with a very high buoyancy move substantially
faster than those having a lower average buoyancy. In this case the
proportionality $w\propto B^{\prime}$, on average, holds fairly well.

\subsection{Scalar dissipation rates}\label{subsec:diss}

\begin{figure}
\begin{centering}
a)\includegraphics[clip,width=0.46\textwidth]{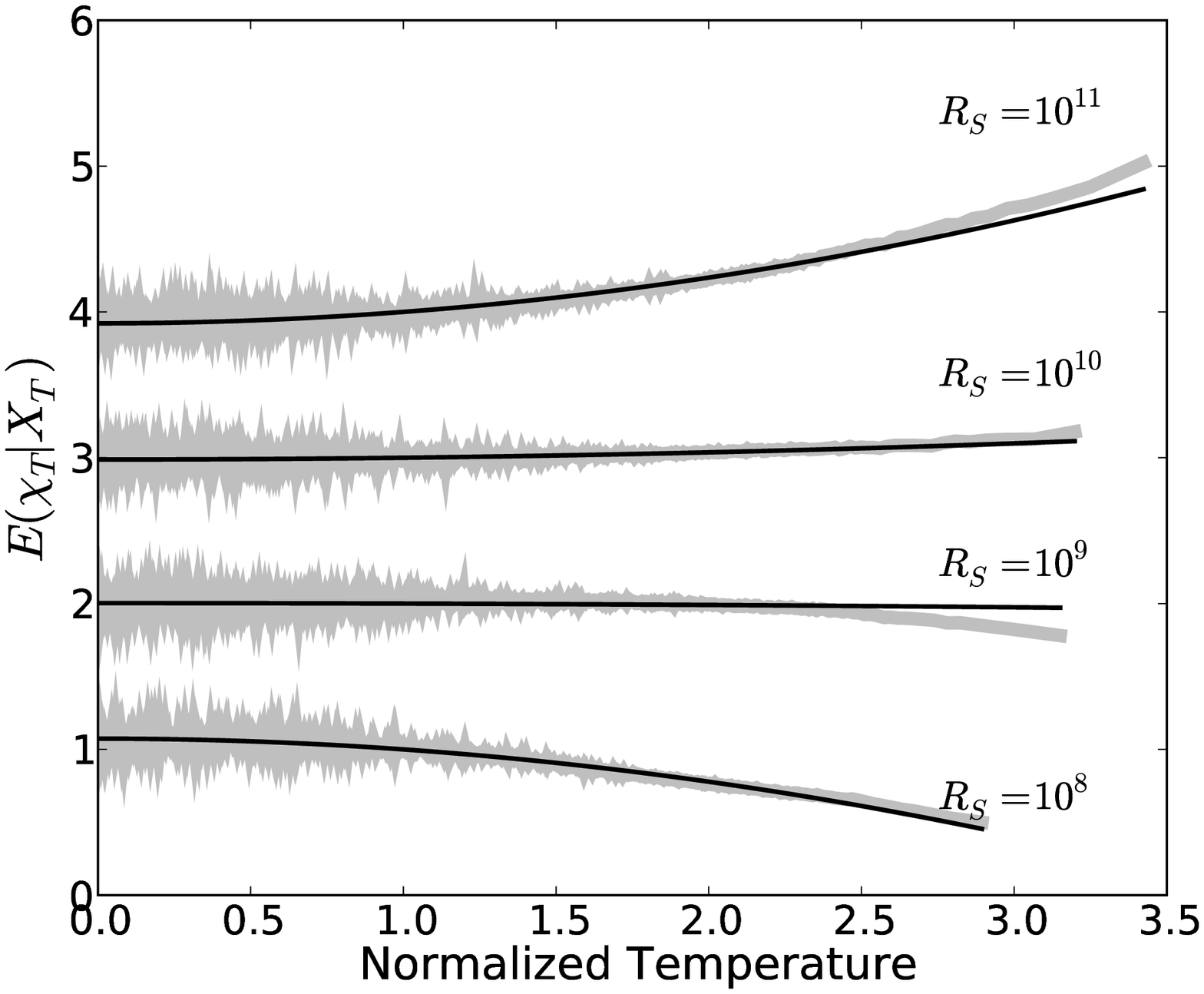}
b)\includegraphics[clip,width=0.46\textwidth]{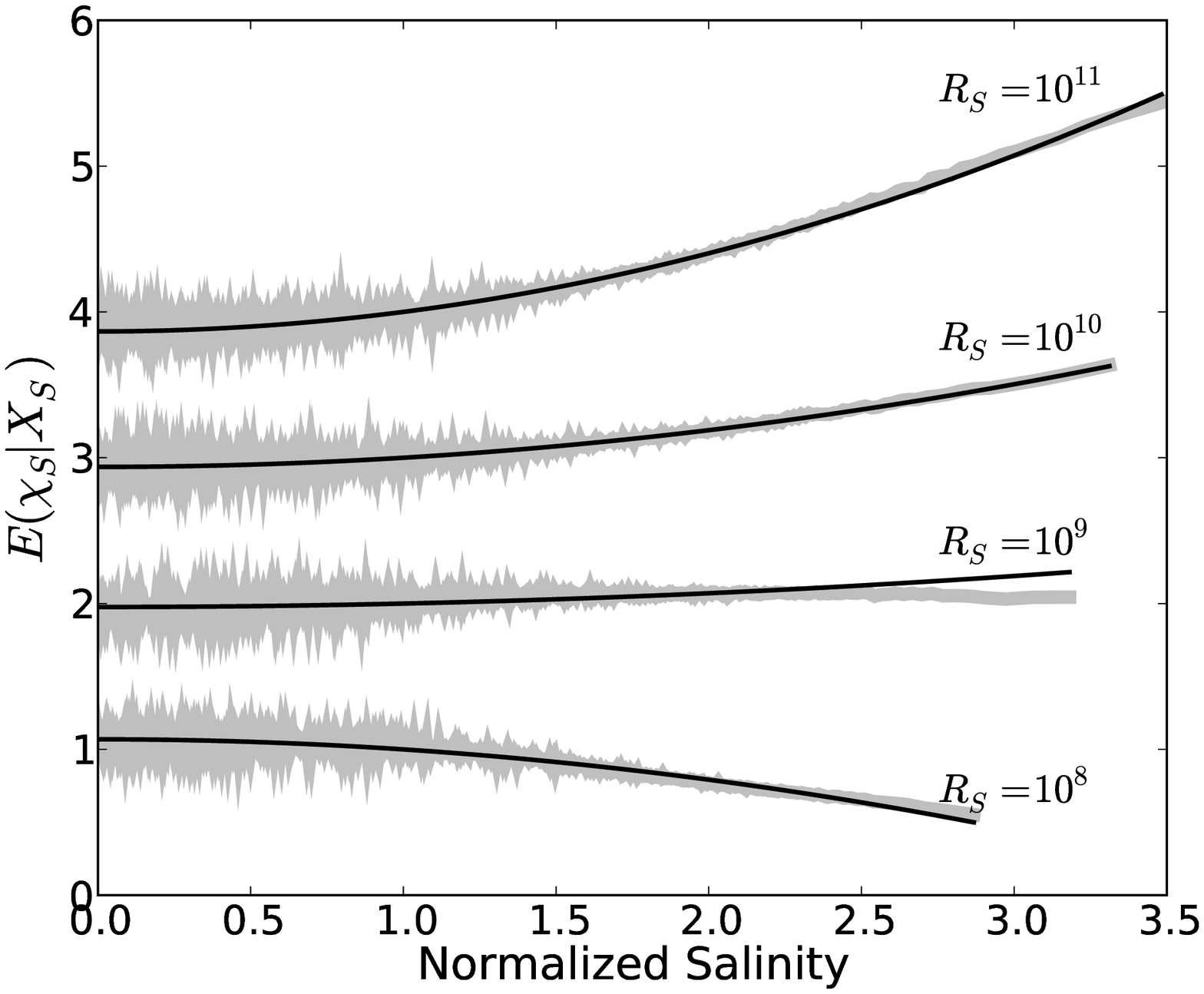}\\
c)\includegraphics[clip,width=0.46\textwidth]{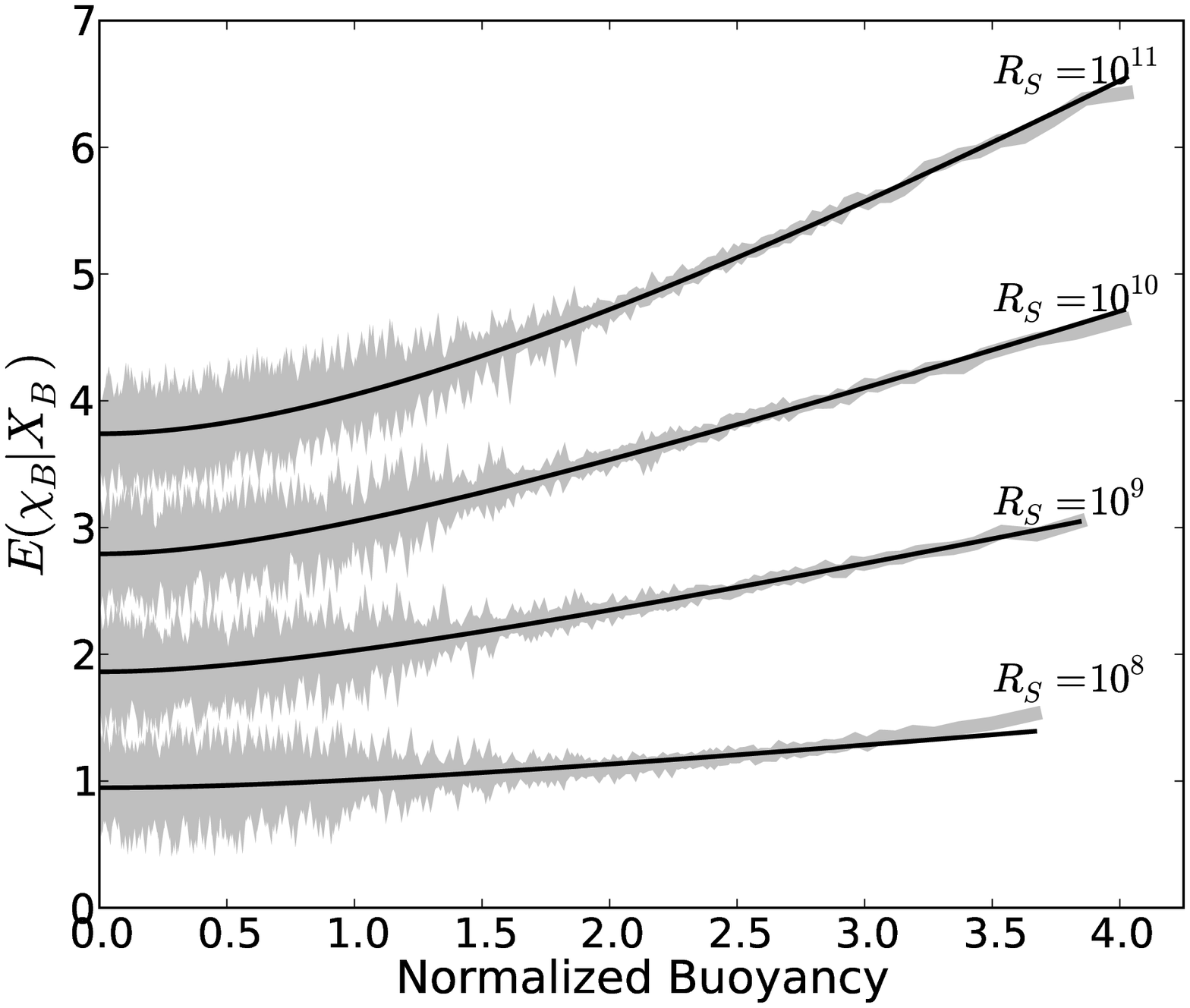}
\par\end{centering}
\caption{\label{fig:Cond-diss}$(a)$ Conditional averaged dissipations of temperature and
  salinity, $E(\chi_{T}|X_{T})$ and $E(\chi_{S}|X_{S})$, for the four simulations at $R_{S}=10^{8}, 10^{9}, 10^{10},
  10^{11}$; data are gathered for one hundred convective times, once
  statistical stationarity has been reached.  For clarity the latter
  three curves are shifted upward by one, two and three units,
  respectively. Overimposed on the numerical data is the
  parabolical fit (\ref{eq:fitchiscal}), where the constant $\gamma$ is 
  determined by a nonlinear least-squares fit. 
  Only the positive normalized fluctuation half-plane is shown. For the negative
  half-plane even symmetry holds. $(c)$ The analogous plot for
  buoyancy dissipation rates, with overimposed  fit
    (\ref{eq:fit-EChiB}).}
\end{figure}

The conditional averages of temperature and salinity
dissipation for the numerical experiments are reported in
Figs. \ref{fig:Cond-diss}(a,b).  To interpret them it is again useful
to take temperature and salinity, to a good approximation, as passive
scalars.  The Kolmogorov-Obukhov-Corrsin scenario
(e.g. \citet{ShraimanSiggia00}) assumes that the dissipation of
variance of a passive scalar is, on average, independent of the
concentration of the scalar itself, which would yield a constant
$E(\mathcal{\chi}|X)$ as a function of $X$. Sinai and Yakhot
\citet{Sinai&Yakhot89} suggested that concentration and dissipation of
a passive scalar may be correlated. They modeled the dissipation as a
parabolic function of concentration and linked this behaviour with
the appearence of non-Gaussian tails.
In our simulations the even parabolic expression
\begin{equation}
 E(\chi|X)=1-\gamma+\gamma X^{2}
\label{eq:fitchiscal}
\end{equation}
fits well the temperature and salinity dissipation data (also in this case the normalization of $X$ and of
$\chi$ reduces the number of free constants in
Eq. (\ref{eq:fitchiscal})).  Conditional dissipation is fairly
constant at $R_{S}=10^{9}$ and smoothly assumes an upward parabolic
shape as the Rayleigh number increases. At low $R_{S}=10^{8}$ we find
a slightly downward shape: high fluctuations of temperature or
salinity are slightly less subject to dissipation than small
ones. This may reflect the particular distribution of
temperature and salinity inside blobs: extreme fluctuations are found
mainly in the regions of low gradients at the core of the blobs,
where they are protected from dissipation.  

Fig. \ref{fig:Cond-diss}(c) shows instead that at all Rayleigh numbers the
dissipation of buoyancy is never independent of buoyancy
itself. To some extent also the dissipation of buoyancy can still
be interpreted as the contribution of a passive scalar. Any buoyancy
fluctuation which is not aggregated into a blob is likely to be
dynamically irrelevant: small buoyancy structures significantly
different from rising or sinking blobs will be strongly damped by
viscosity, and quickly dissipated. Therefore buoyancy in the
background between the blobs is passively transported, and a parabolic
shape of $E(\mathcal{\chi}_{B}|X_{B})$ at moderate values of $X_{B}$
is to be expected. For large values of $X_{B}$ the dissipation is
dominated by the contribution of the blobs.  The detailed distribution
of buoyancy fluctuations within each blob and the distribution of blob
sizes and intensities will determine the form of the tails of
$E(\chi_{B}|X_{B})$ \citet{PH10}. At this stage we limit ourselves to
report that the numerical simulations at high Rayleigh numbers show
remarcably linear tails, and postpone to a future work an in-depth
investigation of the morphology of the blobs.

A simple fitting expression that allows us to join the picture inside
and outside the blobs is 
\begin{equation}
E(\chi_{B}|X_{B})=k+\frac{aX_{B}^{2}}{1+b|X_{B}|}.
\label{eq:fit-EChiB}
\end{equation}
The parameter $k$ weights the Kolmogorean part of the dissipation; $a$
is the coefficient of the quadratic component in the expression of
$E(\chi_{B}|X_{B})$; $a/b$ is the asymptotic slope of the linear
tails. Although the constants $k$, $a$, $b$ are functionally linked
by a normalization constraint, writing this relationship explicitly is
not as straightforward as in the case of the conditional expectation
of the fluxes. Here we prefer to independently fit all three constants
that appear in Eq. (\ref{eq:fit-EChiB}) with a nonlinear least-square
regression to the numerical data. The results are shown in Fig.
\ref{fig:Cond-diss}(c) and show very good agreement.

\subsection{Agreement with the scalar distributions}

Since Eq. (\ref{eq:P(X)}) and its equivalents for temperature and salinity represent
exact expressions for the scalar fluctuation amplitude distributions, computing conditional averages
of the scalar fluxes and dissipations from our experimental data and
substituting them into these equations, leads trivially to an almost
exact overlap with the curves in Fig. \ref{fig:pdfsB} (not
shown). When a simple dependency 
$E(\mathcal{F}|X)/X=X$ is used for the fluxes,
 together with the fits (\ref{eq:fitchiscal}) (for the dissipations of temperature and salinity) 
and (\ref{eq:fit-EChiB}) (for the dissipation of buoyancy), we obtain a very good agreement with the
distributions computed from the numerical simulation data, as shown by
the black lines in Figs. \ref{fig:pdfsB}(a-c).  The only exceptions
are the tails of the buoyancy distribution at low Rayleigh number,
where the agreement is worse due to the deviations from linearity of
$E(\mathcal{F}_{B}|X_{B})/X_{B}$ discussed in
Sec. \ref{subsec:fluxes}.  Note that, when $E(\mathcal{F}|X)/X=X$ is assumed, 
the non-Gaussianity of the tails of the amplitude distributions is controlled by the form of the expected
dissipation: the linear tails of Eq. (\ref{eq:fit-EChiB}) determine
the exponential tails of the distributions of buoyancy fluctuations,
while a dissipation independent of scalar fluctuations (as shown by
temperature and salinity at low $R_S$), leads to Gaussian
distributions.

\section{Conclusions}


In the numerical experiments of fingering convection reported in this
letter we find sharp evidence for exponential non-Gaussian tails in
the buoyancy fluctuation distributions at high Rayleigh number. In
contrast, the statistics of temperature and salinity remain closer to
Gaussianity even when those of buoyancy are already significantly
non-Gaussian.  As shown by using a custom version of a theory by
Yakhot (1989), this observation can be understood in terms of the
different properties of dissipation of a scalar directly creating
vertical accelerations, such as buoyancy, compared to the dissipations
of the individual buoyancy-changing scalars, such as temperature and
salinity.
 
There are some analogies with the phenomenology of
Rayleigh-B\'enard convection, where, in the high Rayleigh number
``hard-turbulence'' regime, temperature fluctuations present
exponential-like tails. In that setting temperature fluctuations are
equivalent to buoyancy, while in fingering convection this role is
played by a linear combination of temperature and salinity.  We
suspect that the faster appearance, in fingering convection, of
non-Gaussian statistics of buoyancy, compared with those of other
active scalars, may apply also to other flows with multiple active
scalars.

The simple conceptual models introduced in Secs. (\ref{subsec:fluxes})
and (\ref{subsec:diss}) highlight the role of coherent dynamical
structures in determining the vertical fluxes and dissipation of
buoyancy fluctuations and consequently the appearance of non-Gaussian
tails in their amplitude distribution. The same statistics for
temperature and salinity are closer to what could be expected of
passive scalars.  The blobs of fingering convection are generated by a
very different mechanism compared to the plumes of Rayleigh-B\'enard
convection and live in a small range of spatial scales where the
effects of molecular diffusion and viscosity are very strong, but
nonlinear terms are just as important.  We have used a simple
threshold in buoyancy to partition the flow and identify the blobs but
a better identification method will be needed to characterize in
detail their structure and to explore their Lagrangian properties,
such as their mean free path, their characteristic life time and the
dynamics of their mutual interactions.

The changes in the shape of the distributions that we observe occur
gradually over a range of Rayleigh numbers spanning three orders of
magnitude. The convective fluxes and other indicators (Table
\ref{tab:statistics}) follow cleanly simple scaling laws, with no
evident breaks.  This suggests that any change in the flow patterns
affecting the distributions is not as dynamically important as, for
example, the changes in the plumes in the analogous transition in
Rayleigh-B\'enard convection. Nevertheless, we are quite confident
that we are not yet observing any sort of ultimate scaling regime of
fingering convection.  In fact, as we have argued in
Sec. \ref{subsec:fluxes}, the vertical convective fluxes can be
modeled in terms of the equilibrium between a blob's buoyancy and a
Stokes drag.  However the scalings of Table \ref{tab:statistics}  imply
that the Reynolds number of an individual blob increases with the
Rayleigh number, and, eventually, it will significantly exceed one. At
that point the dynamics will necessarily change, as the blobs will be
subject to a drag having a nonlinear dependence on velocity. Whether
this will simply mark a change in the slope of the scaling laws and in
the form of the amplitude distributions, or if it will trigger more
dramatic changes, such as the formation of the elusive staircases,
remains to be seen.

\section*{Acknowledgments}

The authors acknowledge support from CASPUR, Roma, Italy, where the
three-dimensional computer simulations where carried out (HPC Standard
Grant 2009).


\begin{thebibliography}{19}
\bibitem{Turner74}J. S. Turner, Ann. Rev. Fluid Mech. \textbf{6},
37 (1974).

\bibitem{Schmitt94}R. W. Schmitt, Ann. Rev. Fluid Mech. \textbf{26},
255 (1994).

\bibitem{Schmitt03}R. W. Schmitt, Progr. Oceanogr. \textbf{56}, 419
(2003).

\bibitem{Schmitt05}R. W. Schmitt, J. R. Ledwell, E. T. Montgomery,
K. L. Polzin, J. M. Toole, Science \textbf{308} 685 (2005).

\bibitem{Baines&Gill69}P. G. Baines and A. E. Gill, J. Fluid Mech.
\textbf{37}, 289 (1969).

\bibitem{Stern75} ch. 11 in: M. E. Stern, {}``Ocean Circulation
Physics'', Academic Press, N.Y. (1975).

\bibitem{Merryfield00}W. J. Merryfield, J. Phys. Oceanogr. \textbf{30}
1046 (2000).

\bibitem{Radko08}T. Radko, J. Fluid Mech. \textbf{609} 59 (2008).

\bibitem{Krishnamurti03}R. Krishnamurti, J. Fluid Mech. \textbf{483}
287 (2003).


\bibitem{Flament02}P. Flament, Progr. Oceanogr. \textbf{54} 493 (2002).

\bibitem{Sinai&Yakhot89}Ya. G. Sinai and V. Yakhot, Phys. Rev. Lett.
\textbf{63} 1962 (1989).

\bibitem{ShraimanSiggia00}B. I. Shraiman and E.D. Siggia, Nature
\textbf{405} 639 (2000).

\bibitem{Yakhot89}V. Yakhot, Phys. Rev. Lett. \textbf{63} 1965 (1989).

\bibitem{Passoni02} G. Passoni, G. Alfonsi, M. Galbiati, Int. J.
Numer. Methods Fluids \textbf{38} 1069 (2002).

\bibitem{Parodi04} A. Parodi, J. von Hardenberg, G. Passoni, A. Provenzale,
E.A. Spiegel, Phys. Rev. Lett. \textbf{92} 194503 (2004).

\bibitem{Parodi08} J. von Hardenberg, A. Parodi, G. Passoni, A. Provenzale,
E.A. Spiegel, Phys. Lett. A \textbf{372} 2223 (2008).

\bibitem{Calzavarini06}E. Calzavarini et al., Phys. Rev. E \textbf{73}
035301 (2006).

\bibitem{Castaing89} B. Castaing et al., J Fluid Mech. \textbf{204}
1 (1989).

\bibitem{PH10} F. Paparella and J. von Hardenberg, in: A. M. Greco,
  S. Rionero, T. Ruggeri (Eds.) Proceedings of WASCOM 2009, World
  Scientific, in press (2010).
\end{thebibliography}
\end{document}